# Correlative chemical and structural nanocharacterization of a pseudo-binary $0.75Bi(Fe_{0.97}Ti_{0.03})O_3$–$0.25BaTiO_3$ ceramic

Shane J. McCartan[1,2], Ilkan Calisir[3], Gary W. Paterson[1], Robert W. H. Webster[1], Thomas A. Macgregor[1], David A. Hall[3], Ian MacLaren[1]

[1]*School of Physics and Astronomy, University of Glasgow, Glasgow G12 8QQ, UK.*

[2]*School of Mathematics and Physics, Queen's University of Belfast, Belfast, BT7 1NN, UK.*

[3]*Department of Materials, University of Manchester, Manchester, M13 9PL, UK.*

## Abstract

Fast-cooling after sintering or annealing of $BiFeO_3$-$BaTiO_3$ mixed oxide ceramics yields core-shell structures that give excellent functional properties, but their precise phase assemblage and nanostructure remains an open question. By comparing conventional electron energy loss spectroscopy (EELS) with scanning precession electron diffraction (SPED) mapping using a direct electron detector, we correlate chemical composition with the presence or absence of octahedral tilting and with changes in lattice parameters. This reveals that some grains have a 3-phase assemblage of a $BaTiO_3$-rich pseudocubic shell; a $BiFeO_3$-rich outer core with octahedral tilting consistent with an *R3c* structure; and an inner core richer in Ba and even poorer in Ti, which seems to show a pseudocubic structure of slightly smaller lattice parameter than the shell region. This last structure has not been previously identified in these materials, but the composition and structure fit with previous studies. These inner cores are likely to be non-polar and play no part in the ferroelectric properties. Nevertheless, the combination of EELS and SPED clearly provides a novel way to examine heterogeneous microstructures with

Corresponding author: ian.maclaren@glasgow.ac.uk

high spatial resolution, thus revealing the presence of phases that may be too subtle to detect with more conventional techniques.

**Introduction**

Mixed oxide electroceramics are of great interest due to their unique functional properties and therefore, the abundance of technological applications such as piezoelectric sensors, transducers and actuators(1–4). It is well known that tuning material properties, such as stoichiometry, elemental composition or epitaxial strain, can enhance piezoelectric properties(5,6). In particular, these properties are greatly enhanced around the morphotropic phase boundary – a region in the phase diagram in which two structurally competing phases coexist within a narrow range of stoichiometry.

Currently, the most commercially used piezoceramic is the inorganic compound $Pb(Zr_xTi_{1-x})O_3$ (0<x<1) (PZT), due to its excellent functional properties(7). One drawback to this compound, however, is its environmental toxicity due to its lead content. The threat of incoming EU regulations regarding the use of lead has created a surge of research dedicated to finding a suitable, lead-free replacement material with equivalent or superior properties(8). One candidate is the pseudo-binary solid solution, $BiFeO_3$-$BaTiO_3$ (BF-BT). $BiFeO_3$ itself is a promising candidate for many devices, due to its high remanent polarisation and high Curie temperature(9). However, its formation generally yields impurity phases(10,11) and the high coercive field and high leakage current results in unsaturated hysteresis loops being reported(9,12). Many attempts have been made to improve the properties of $BiFeO_3$ based ceramics by doping or using different synthesis methods(13–16), although this can result in unexpected defect formation(17–19). However, addition of $BaTiO_3$ into the solid solution has been reported to stabilise the $BiFeO_3$ perovskite structure and improve the electrical properties(20–22).

The BF-BT solid solution exists in a rhombohedral phase for BiFeO$_3$ contents $\gtrless 0.70$. Below this threshold, the material adopts a cubic structure, before transforming to a tetragonal phase for BiFeO$_3$ content $\lessgtr 0.1$(20,21). The use of certain types of dopants in BF-BT has been shown to cause chemical segregation resulting in core-shell microstructures, a process attributed to an electronegativity difference of the dopant species driving immiscibility during cooling from the sintering temperature(23,24). Initially seen as undesirable due to the reduction in the strength of functional properties, Murakami *et al.*, showed that quenching after sintering or annealing could eliminate chemical heterogeneity and was accompanied by increased electrostrain and polarisation, suggesting that quenching provided potential for improved functional properties(23). More intriguingly, it has been shown that donor dopants promote chemical heterogeneity, while isovalent dopants promote solubility(25). Therefore, the consensus derived from a large number of studies on BF-BT, is that the structure and properties are highly sensitive to the doping species, doping quantity and heat treatment, suggesting that there may be more than one route to tune the desired functionality of these ceramics. In a further publication, Calisir *et al.*(26) propose a new avenue of exploration of piezoelectric properties, in which they exploit deliberate chemical segregation and the creation of a core-shell microstructure. By creating both chemically and structurally inhomogeneous ceramics, they report enhanced dielectric energy storage density(25) with slow-cooling and improved piezoelectric properties with fast-cooling(26) in core-shell structured BiFeO$_3$-BaTiO$_3$. In disagreement with the findings of Murakami *et al.*(23) that quenching reduces heterogeneity, it was found that the core-shell microstructure was maintained with quenching accompanied by a dramatic enhancement of the ferroelectric and piezoelectric properties with respect to furnace cooled samples.

With a new vision to create chemically and structurally inhomogeneous materials in order to generate enhanced piezoelectric properties, it is necessary to have methods to characterise this inhomogeneity at the requisite length scale (tens of nanometres or less). Distinguishing these details using X-ray diffraction or scanning electron microscopy (SEM) techniques such as X-ray mapping or electron backscatter diffraction (EBSD) is not realistic due to the lack of spatial resolution and the relatively subtle differences between the phases involved(27,28). To make progress in correlating structure and chemistry, here we study thin sections using a combination of the high spatial resolution techniques of electron diffraction and EELS in the transmission electron microscope/scanning transmission electron microscope (TEM/STEM).

Specifically, the study of nanostructured mixed phases is an ideal application for SPED. First demonstrated in 1994(29), precession electron diffraction is a technique in which a convergent electron beam is focused on the sample, tilted away from the optic axis, and precessed about the optic axis, whilst the diffraction pattern is recorded. This averages out dynamical diffraction effects and the resulting intensity distributions of the integrated diffraction patterns are considered pseudo-kinematical. More recently, this was turned into a scanning technique in which the beam is simultaneously precessed and rastered over the sample(30), thus making it one variant of a scanned diffraction or 4-dimensional STEM (4D STEM) technique(31–33) with resolutions down to a few nanometres, making it perfect for mapping core-shell structures.

It is well-known that octahedral tilting of perovskites causes the appearance of extra reflections in electron diffraction patterns which reveal the nature of the tilting pattern(34). Previously, $Bi(Fe_{1-x}Ti_x)O_3$-$BaTiO_3$ was found to be a mixture of a rhombohedral *R3c* perovskite ($BiFeO_3$-like) containing substantial antiphase octahedral tilting and a primitive pseudocubic perovskite (maybe similar to $BaTiO_3$) with untilted oxygen octahedra(26). Given that the similarity of $BaTiO_3$ and $BiFeO_3$ structures (~1% distortion from cubic perovskite)(9,35) would make most equivalent zone axes diffraction patterns indistinguishable, the presence of superlattice reflections at the ½{*ooo*} positions (where *o* stands for an odd number) arising from antiphase octahedral tilting(36), was utilised to spatially map the crystal structure. The lowest index zone axes in which these reflections appear are the <110> and <112> zone axes of the primitive perovskite cell. Woodward and Reaney(34) analysed the appearance of superlattice reflections in <110>$_{primitive}$ zone axes for perovskites with different tilting systems and showed that these only appear in 6 of the 12 possible <110>$_{primitive}$ directions for $a^-a^-a^-$ structures like BiFeO$_3$. Thus, using the presence or absence of such reflections in a <110>$_{primitive}$ direction is not reliable, as their absence could also indicate a domain with a different crystallographic orientation. In contrast, all 24 distinct <112>$_{primitive}$ directions contain allowed ½{*ooo*} reflections (a detailed analysis using calculated diffraction patterns is included in the *Supplemental Materials* Figure S1 and Table S1), making these the optimal zone axes for diffraction experiments to distinguish octahedrally-tilted and untilted perovskites.

Once acquired, SPED data can be visualised using a Virtual Dark Field (VDF) approach, where an image is generated using the intensity in a single diffraction spot(37) or multiple diffraction spots from the same phase(38), allowing the separation of different crystals, phases or domain orientations in a crystal. Additionally, mapping the spot positions allows the determination of strain(39–42). The first of these, however, would be difficult for weak antiphase tilting spots using the conventional SPED detector (a fast CCD camera), as the weak spots would be lost in the relatively high background noise level. In a recent advance, an electron counting pixelated detector (Merlin for EM with Medipix 3 chip) was integrated with a SPED system(43). This allows SPED data collection at a suitable signal to noise ratio to extract the intensity of the superlattice spots.

This paper describes a study of air-quenched Bi(Fe$_{0.97}$Ti$_{0.03}$)O$_3$-BaTiO$_3$ ceramics in a TEM/STEM system using EELS chemical mapping coupled with analysis of weak superlattice spots in SPED patterns that reveal octahedral tilting. This reveals the core-shell structure in unprecedented detail, and even leads to the discovery of an additional unexpected phase in the ceramic.

**Experimental Procedure**

Samples were synthesised by the solid state reaction method based on the chemical formula of 0.75(3% Ti-doped $BiFeO_3$)−0.25$BaTiO_3$. More details are given by Calisir *et al.*(26). Samples for TEM/STEM analysis were prepared by a focused ion beam (FIB) liftout process using a Helios xenon plasma FIB (Thermo Fisher Inc., Hillsboro, OR).

All TEM and STEM data were collected on a JEOL ARM200F (JEOL Ltd., Akishima, Japan) with a cold field emission source using an accelerating voltage of 200 kV. EELS measurements were acquired in STEM mode with a Gatan GIF Quantum ER electron energy loss spectrometer (Gatan Inc., Pleasanton, CA) using a beam convergence angle of 29 mrad and spectrometer collection angle of 36 mrad. The spectra were acquired with an energy dispersion of 0.5 eV/channel over the range of 1024 eV. This encompassed major edges of four of the constituent elements in the composition. Any discernible bismuth edge lay well outside the acquired energy range and therefore bismuth could not be mapped simultaneously, however the acquired high-angle annular dark field (HAADF) image can be taken as a measure of bismuth presence, as contrast scales with atomic number in HAADF imaging and bismuth is the heaviest constituent element. Relating to the EELS mapping, a separate EELS map of the same area was obtained over the range of 1812-3848 eV with a dispersion of 1 eV/channel to map bismuth directly using the $M_{4,5}$ edge (see *Supplementary Information* Figure S3).

EELS data were processed in Digital Micrograph (Gatan Inc., Pleasanton, CA) by firstly removing any X-rays spikes from the data due to X-ray generation from scattering. The spectra were then aligned using the zero-loss peak to account for movements of the spectrum on the detector during acquisition (either from energy drift or scanning effects). Multivariate statistical analysis was then performed on the data using a specially designed plugin in Digital

Micrograph to separate the real signal from random noise(44). The elemental composition was then quantified using the *Elemental Quantification* plugin in Digital Micrograph and chemical maps of the constituent elements that lay within the acquired energy range were generated using the Ti and Fe $L_{2,3}$ edges, the Ba $M_{4,5}$ edge and the O K edge.

The EELS spectra from the different regions were quantified to produce relative contents of Bi, Ba, Ti and Fe. The Ti and Fe contents were taken straight from the standardless quantification using the formula $(Bi_{1-x}Ba_x)(Fe_{1-y}Ti_y)O_3$ and assuming that the Fe and Ti amounts always sum to 1, and thereby reporting *y* and *1-y*.

It was slightly more involved for Ba and Bi as these are in separated spectrum images of the same area. To account for this, spectra from the same integrated areas were created for both scans and the integrated intensity in the edge determined for each of the shell, outer core and inner core regions, was divided by the zero loss intensity to produce:

$$\frac{I}{I_0} = N\sigma$$

for each element, where *I* is the intensity in the edge, $I_0$ is the intensity in the zero loss peak, *N* is the number of atoms per unit area in the path of the beam, and $\sigma$ is the interaction cross section. Whilst we do not know the cross-section values (and estimates for M edges are likely to be inaccurate(45)), we can determine an approximate composition in a 2-component system where the two concentrations have to add up to a constant using a simple simultaneous equations approach. This was already found to give a close approximation to results found by determining absolute cross sections from standards for Si-Ge alloys(46). In this case, the problem is overdetermined as we have three regions to work from, not just two, as in the Si-Ge. Nevertheless, the *k*-factor relating the cross sections for Bi and Ba can be estimated and then adjusted until the residuals in the analysis are minimised. In our case, with a window width of 400 eV starting at 2600 eV for the Bi-$M_{4,5}$ edge (to minimise noise) and a window

width of 50 eV starting at 781 eV for the Ba-$M_{4,5}$ edge, the *k*-factor was found to be about 2.1, i.e.:

$$\left[\frac{I}{I_0}\right]_{Bi} + 2.1 \left[\frac{I}{I_0}\right]_{Ba} \approx Constant$$

for the three regions. Therefore, the Ba and Bi fractions could be determined as:

$$x = \frac{2.1\left[\frac{I}{I_0}\right]_{Ba}}{\left[\frac{I}{I_0}\right]_{Bi} + 2.1\left[\frac{I}{I_0}\right]_{Ba}} \quad \text{and} \quad 1 - x = \frac{\left[\frac{I}{I_0}\right]_{Bi}}{\left[\frac{I}{I_0}\right]_{Bi} + 2.1\left[\frac{I}{I_0}\right]_{Ba}}$$

SPED data was collected using the JEOL ARM200F noted above with an acceleration voltage of 200kV, a NanoMEGAS Topspin software system (with a prototype system to use the MerlinEM direct electron detector [Quantum Detectors Ltd., Harwell, Oxfordshire, UK] for the diffraction pattern acquisition) and Digistar hardware (NanoMEGAS SPRL, Brussels, Belgium), using a beam precession angle of 0.5° and a 10μm condenser aperture to obtain a low convergence angle and well-separated diffraction spots. Comparison diffraction patterns from the same region were also recorded using the standard Stingray camera (NanoMEGAS SPRL, Brussels, Belgium) focused on the small viewing screen.

The SPED data was processed to generate structural maps using python scripts derived from the *fpd* library, which is detailed by Paterson *et al.*(38). Functions in the *fpd* library were used for identifying the central beam spot and then cross correlation and threshold discriminators were used to identify spots with similar characteristics, i.e. diffracted spots. These were then filtered to only include any which had a Friedel counterpart to create a synthetic lattice. This included all primitive lattice spots, excluding any superlattice spots (achieved through intensity threshold discrimination). This synthetic lattice was then used to generate a mask with an array of virtual apertures at the primitive spot positions, from which a VDF image could be generated and normalised to the total counts per real space pixel. The mask was then displaced a certain amount, such that the aperture array then encompassed the superlattice reflections, from which a second VDF image was generated. The strain mapping feature in the TopSpin software was used to map changes in lattice parameters across a grain of interest (based on work by Darbal *et al.*(39–42)). Changes in lattice parameter were mapped in the sample in two orthogonal directions and correlated to changes in octahedral tilting and chemistry.

## Results

The comparison between the two detectors for the same diffraction pattern from the same crystal shows a huge improvement in the diffraction data for the direct electron counting detector, as illustrated in Figure 1. Specifically, the superlattice reflections are now clearly distinguished from the noise, allowing good prospects for mapping their appearance as a function of scan position. This detector was therefore used in the remainder of this work.

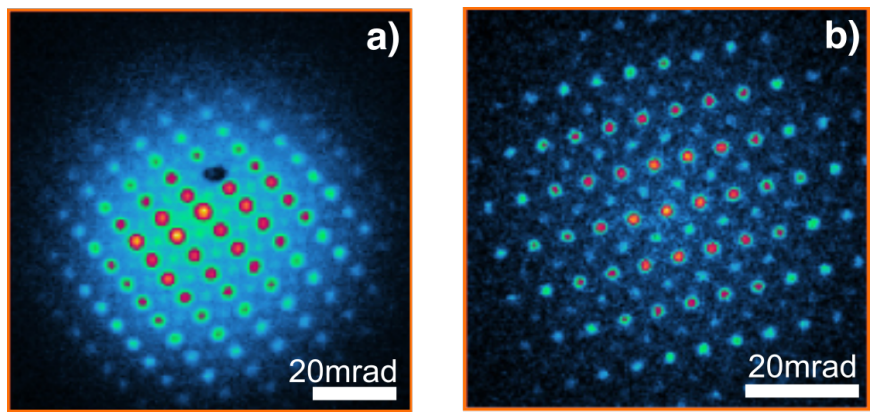

*Figure 1 – Perovskite <110>$_{primitive}$ SPED patterns obtained using a) the optically coupled Stingray camera; and b) the Medipix 3 direct electron detector. In both cases, the area is an area of perovskite with octahedral tilting and superlattice spots viewed along <110>.*

Figure 2 shows chemical maps generated from the EELS data. Clearly, the structure shows chemical heterogeneity within grains in the form of phase segregation of the two perovskite end members ($BiFeO_3$ and $BaTiO_3$) into a core-shell structure, consistent with previous reports on this material(26). It is clear that the Ba and Ti maps correlate closely with one another and mainly show regions at the outside of grains rich in these elements. These elements are generally anti-correlated with both the Fe map and the HAADF image and the Bi map (see *Supplemental Materials* Figure S3). The O signal seems to be relatively homogeneous through the grains. This is consistent with $BaTiO_3$-rich shells and $BiFeO_3$-rich cores, which has been attributed mainly to the $Ti^{4+}$ donor doping in place of $Fe^{3+}$ causing differences in electronegativity and hence some degree of immiscibility between $BaTiO_3$ and $BiFeO_3$, inhibiting them from mixing in a solid solution. Another region worth noting is the central core of the lower grain on the right-hand side. In the inner core region, there appears to be a correlation (increased intensity) between the Ba and Fe maps, which both appear to be anticorrelated with the HAADF image and Ti map (decreased intensity), causing an orange-coloured area in the centre of the core of this grain in Figure 2f. There are also orange regions in the grain to the left and the grain above, suggesting that this may be common to many if not all grains (especially as slicing a lamella from the bulk will not always cut directly through the centre of grain). This suggests that the phase formed in these regions is distinct from the surrounding core or shell regions and appears to more barium ferrite rich than other areas. It may also be noted that there are some bright regions at grain boundaries in the HAADF image Figure 2d. These are also apparent in the bismuth EELS map of Figure S3a and they correspond to the bright areas of the HAADF image in Figure S3b. It is likely these are excess $Bi_2O_3$ regions, resulting from the excess Bi used in sintering to compensate bismuth loss due to vaporization.

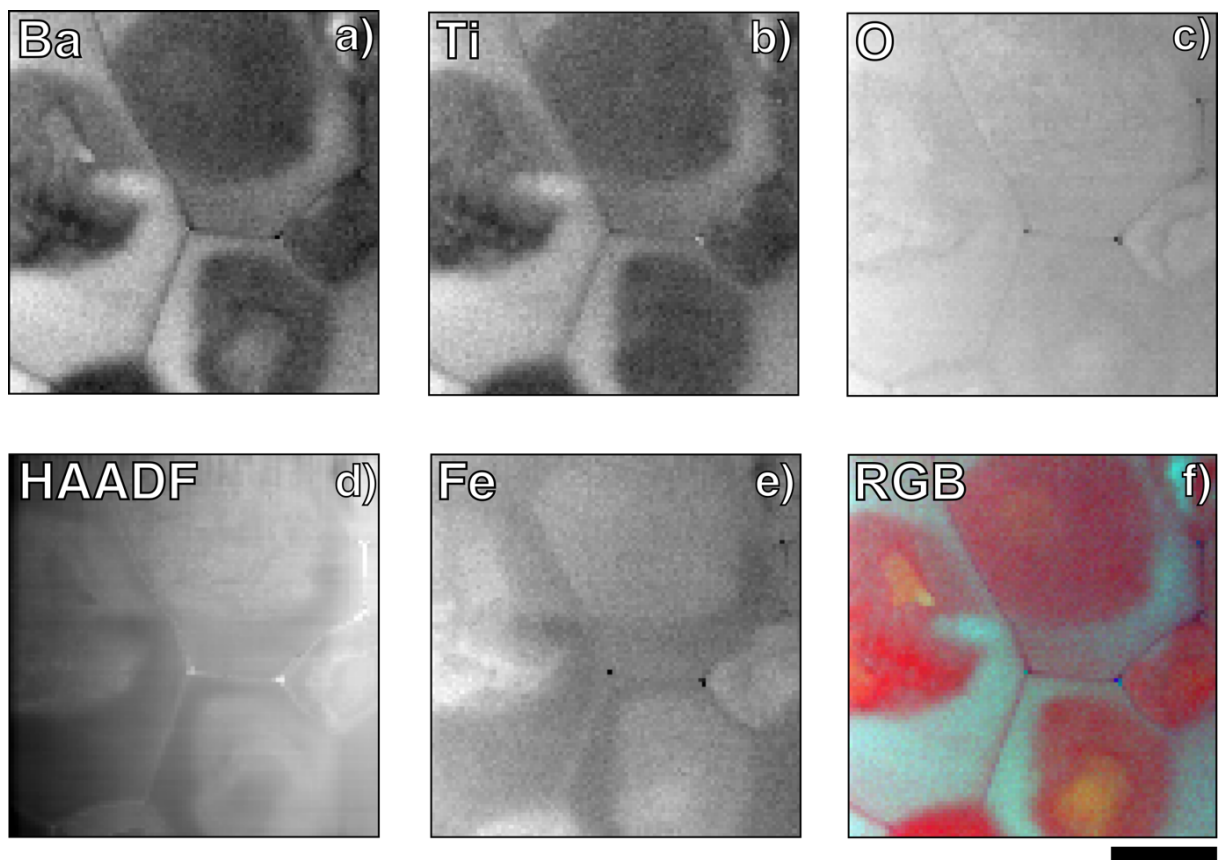

*Figure 2–Elemental maps from EELS and the HAADF signal of a multi-grained region of the specimen. (a) Ba M signal, (b) Ti L signal, (c) O K signal, (d) HAADF image, (e) Fe L signal and (f) a composite RGB map consisting of Fe (red), Ti (green) and Ba (blue). The scale bar represents 500nm.*

Table 1 summarises quantification results from the EELS data. The first thing to note is that there are no areas of pure $BiFeO_3$ or $BaTiO_3$; every area is a solid solution. Nevertheless, the shell is around 50-60% $BaTiO_3$, which puts this into the pseudocubic region of the $BaTiO_3$-$BiFeO_3$ phase diagram(20,21). In the outer core, the Ba content is very low and these regions are more likely to be in the rhombohedral region of the phase diagram with a structure closer to that of $BiFeO_3$. Finally, there is a region where Ba is at an intermediate composition, but Ti is at its lowest and Fe is relatively high. This is clearly not on the $BiFeO_3$-$BaTiO_3$ tie line, but is maybe heading towards a $BaFeO_3$ composition away from this tie line. It should be noted that whilst the Ba:Bi ratios are plausible for a starting composition of 25%$BaTiO_3$:75%$BiFeO_3$, the Ti contents seem too high as any sensible weighted average would yield something larger than 0.25, and there is probably an error in the calculation of the cross sections used in quantification resulting in a systematic overestimate of Ti content. Nevertheless, even if the absolute magnitude of these proportions could be subject to some significant systematic errors (at least 0.05 and possibly 0.1), the general trends are still clear.

|  | Formula fractions of each element | | | |
| --- | --- | --- | --- | --- |
| Grain area | Ba $x$ | Bi $1-x$ | Ti $y$ | Fe $1-y$ |
| Shell | 0.54 | 0.46 | 0.60 | 0.40 |
| Outer core | 0.19 | 0.81 | 0.34 | 0.66 |
| Inner core | 0.38 | 0.62 | 0.22 | 0.78 |

*Table 1: Cation compositions calculated for the shell, outer core and inner core regions. As explained more fully in the text, the absolute quantities should be treated with care, but the trends are still illustrative of what is happening in the segregation process.*

In order to map the local crystal structure of the outer core, shell and inner core phases, SPED was used and exemplar diffraction patterns from the shell and the core are shown in Figures 3a, and b. Weak superlattice reflections are clearly seen in 3b, exactly as expected for antiphase tilting. The masks used for the software processing of the intensity in the main primitive perovskite and the superlattice spots are shown in Figure 3c.

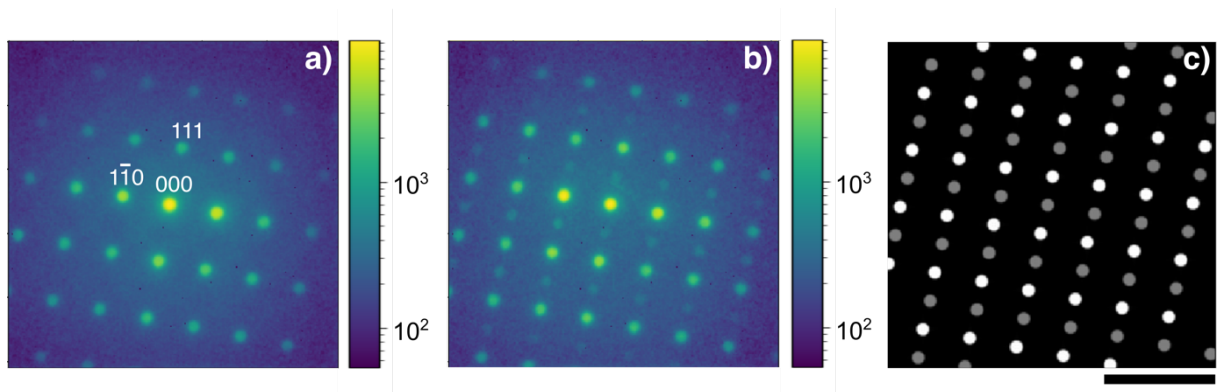

*Figure 3 – <112>$_{pc}$ SPED patterns obtained from the shell (BaTiO$_3$ rich) region (a) and from an outer core region (BiFeO$_3$ rich) of the grain (b). (c) shows the lattice of synthetic apertures used to generate the VDF images – grey lattice corresponds to apertures for primitive spots, while the white lattice corresponds to apertures for superlattice spots. The scale bar represents 20 mrad.*

Figure 4 shows virtual dark field images generated from a scan across the area containing the three-phase grain in the bottom right corner of the EELS maps presented in Figure 2 (scan is rotated in data presented hereafter). Figure 4a shows a VDF image for the primitive lattice spots, and unsurprisingly highlights the whole grain, although there is a slight dip in intensity at the top of the grain and in the inner core region.  Figure 4b shows just the intensity in the superlattice spots and highlights most of the core, although the inner core is notably missing from this image. The dark and light regions in this image correspond fairly well with the dark and light regions in the HAADF image of Figure 4c, although the contrast in that image is much less.  Comparing these images to the EELS map of Figure 4d (in the same 3-colour scheme used in Figure 2f), it is clear that the area with strong superlattice spots is that which is most BiFeO$_3$-rich, whereas the BaTiO$_3$-rich shell and the barium ferrite rich inner core do not have these superlattice spots and thus do not have antiphase tilting.  It should be noted that the diagonal stripes in Figures 4a and 4b and the horizontal stripes in Figures 4c and 4d are purely scan artefacts and do not convey any information about real structures in the grain.  Again, some bright contrast is seen on the grain boundaries and triple points, especially on the right, corresponding to Bi$_2$O$_3$ segregation.  There is also a bright particle in the lower right of Figure 4c that appears as black in the EELS map of Figure 4d, this is probably a piece of dust or contamination on the surface of the sample that was not present when the data shown in Figures 4a and 4b were recorded.

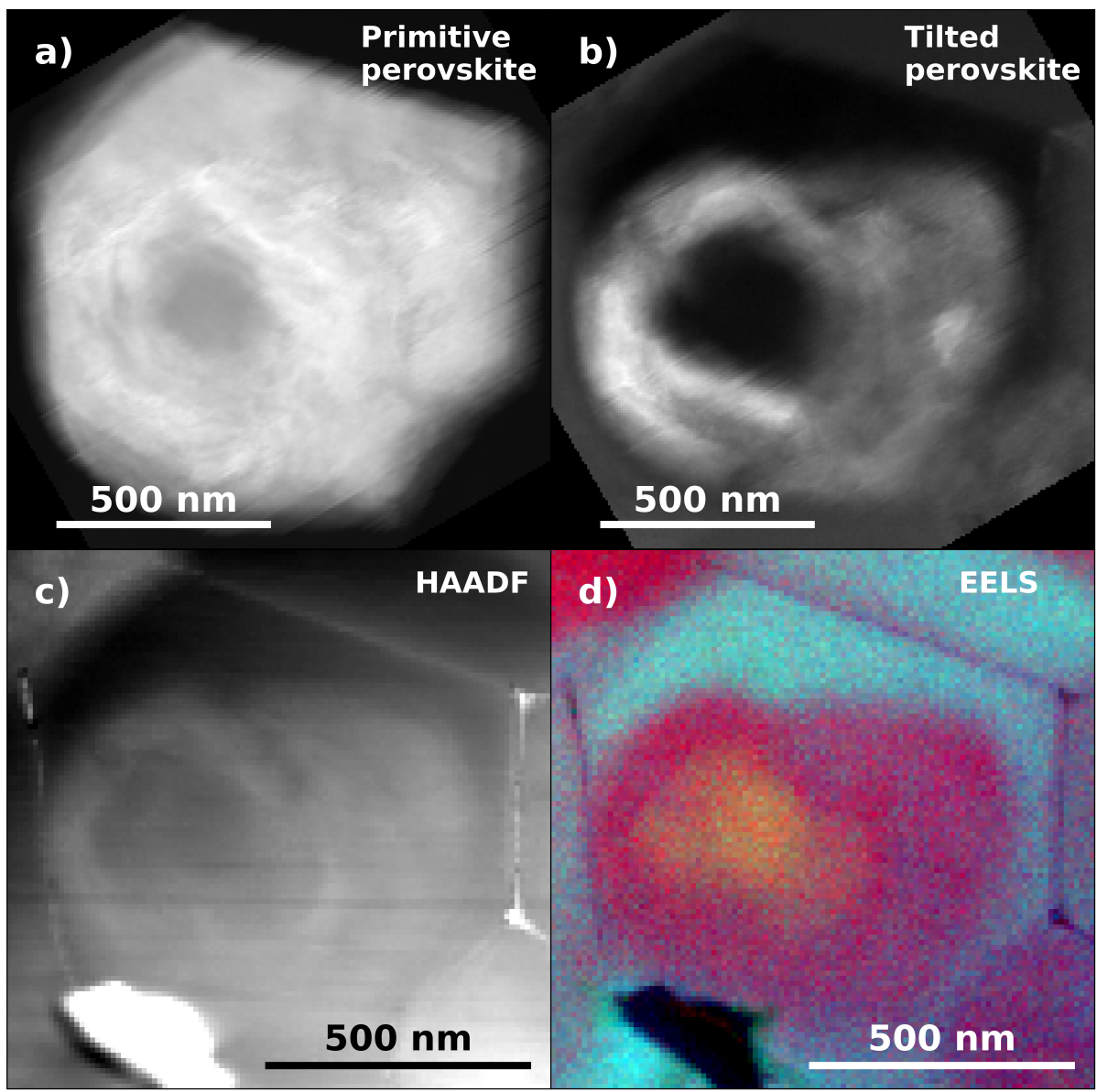

*Figure 4 – Correlation of the structural and chemical mapping in one grain imaged along <112>$_{pc}$: VDF images generated from contributions of the primitive lattice spots (a) and the superlattice spots (b); (c) shows a HAADF image of the grain of interest (which is indicative of Bi content); and (d) an RGB composite EELS map using the same colour scheme as in Figure 3.*

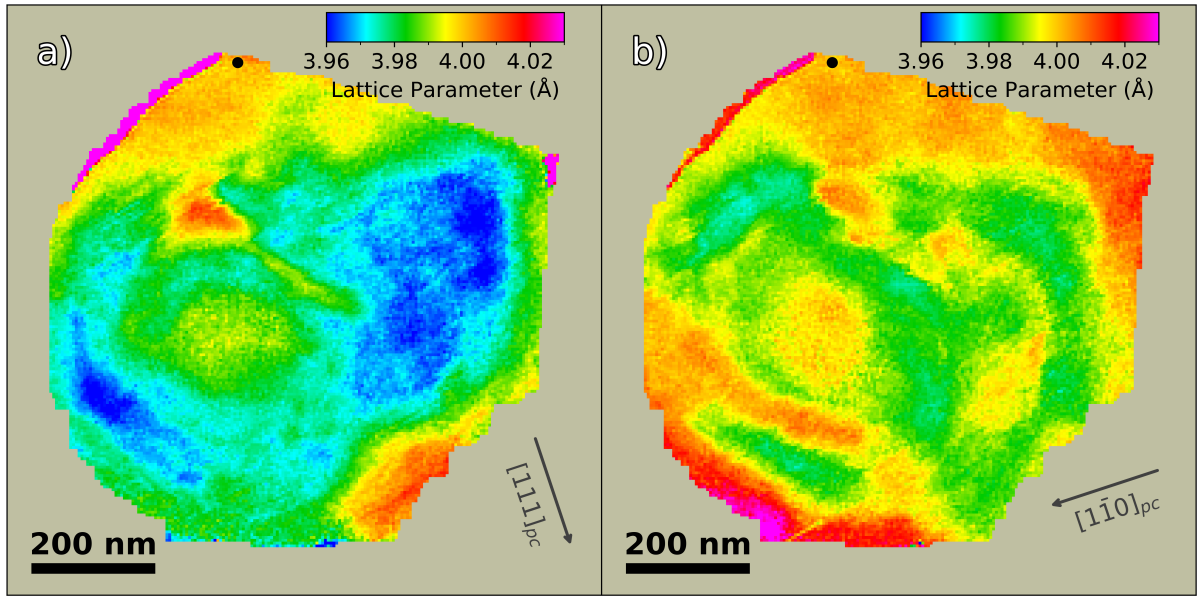

*Figure 5 – Lattice parameter maps with respect to a barium titanate rich region (denoted by the black dot) along a) $[111]_{pc}$ and b) $[1\bar{1}0]_{pc}$ directions.*

In order to map the lattice parameters, the inbuilt strain mapping function in the ASTAR software was used on a dataset from the same area, which requires a reference point to be selected. The lattice parameter was mapped relative to a chosen point within the grain. A BaTiO$_3$ rich region was chosen as a reference point and is indicated by a black dot in both figures. The reference lattice parameter for the black dot was set as 4Å as this is about the lattice parameter typically quoted for pseudocubic phases in BiFeO$_3$-BaTiO$_3$ solid solutions(23) (technically, this is the pseudocubic lattice parameter, $a_{pc}$, calculated from the data, whereas the diffraction pattern spacings in the two orthogonal directions will actually depend on $\sqrt{3}/a_{pc}$ and $\sqrt{2}/a_{pc}$). Figures 5a and b map the lattice parameter along the orthogonal $[111]_{pc}$ and $[1\bar{1}0]_{pc}$ directions, indicated with arrows. The $[111]_{pc}$ direction was chosen as this denotes the polar direction of bismuth ferrite and there may therefore be an associated strain/elongation along this direction. Alternately, other regions of bismuth ferrite may see a contraction along this direction and an expansion along another direction, such as a $[1\bar{1}0]_{pc}$ direction. If one compares Figure 5 with Figure 4, the areas giving strong superlattice contrast tend to have small lattice parameters (green or blue), whereas the shell tends to be mostly larger lattice parameters (reddish or yellow), and the inner core is also yellow or orange suggesting that it is also a larger lattice parameter structure. It is, however, the case that not all parts of the $BiFeO_3$-like region have the same lattice parameters and redder parts are seen where the lattice parameter in one or other direction is around or just over 4Å. These may be domains where the **c**-axis lines up along one specific direction. It is notable that two of the redder parts in Figure 5b in the outer core show very blue contrast in Figure 5a, suggesting just such a distortion in the plane of the diffraction pattern where it is most detectable.

## Discussion

It is clear from the data presented that the core-shell structure in the quenched samples is a 3-phase one, not the two-phase structure previously posited(25,26). The three phases will be discussed in turn in the forthcoming paragraphs.

The shell has a composition where about 60% of B site atoms appear to be Ti. This area is also enriched in Ba, although it does not appear that the A sites are quite as rich in Ba as the B sites are in Ti. This correlates with a reduction in intensity in HAADF images, which should be brightest in Bi rich areas because of the strong atomic-number contrast in HAADF imaging and the fact that Bi is by far the heaviest element in this ceramic. It also correlates with a reduction in the EELS signal to background intensity at the Bi $M_{4,5}$ edge, even if that had to be recorded separately as it appears in a very different energy range to the other element edges (see *Supplementary Information* Figure S2). It is therefore evident that this is the area of the grain that is richest in $BaTiO_3$. Moreover, whilst the detailed analysis was performed on one grain, larger area EELS maps show similar trends across many grains in the ceramic. Using precession electron diffraction, this area consistently did not show ½{$ooo$} reflections, which are a sign of antiphase tilting(23,26,34) – even careful examination of the raw data focusing on those positions found no evidence for these. This either suggests that the structure is an untilted pseudocubic perovskite, as has been suggested in previous reports on the more $BaTiO_3$-rich phases in this system(23,47), or a rhombohedral phase with very small tilting such that the intensity in the superlattice spots is undetectably small(26). Unsurprisingly, these areas are generally those with the largest lattice parameter, as was already expected from previous publications on this system(23,26).

The outer core was, as expected, rich in $BiFeO_3$, although Ti and Ba are still present at lower concentrations. Nevertheless, the precession electron diffraction clearly demonstrates strong ½{*ooo*} reflections throughout this area, giving clear evidence that this area has formed the rhombohedral $BiFeO_3$ structure expected. This area also shows some of the smallest lattice parameters, but with different contractions in different areas, which would agree with the observation of a complex domain structure for this area using conventional TEM (*see Supplementary Information* Figure S4).

It should be noted that the nature of the core-shell structure is crucial in determining the functional properties of the system. For instance, when this material was compared to material that was slow-cooled, the boundary of the outer-core and the shell regions seemed less well defined in the slow-cooled material. On the other hand, when the material is quenched, the clear boundary between the shell and the outer core and distinct chemistry and crystallography of the two as shown in the results above can be expected to lead to long range polarisation ordering in the $BiFeO_3$-rich parts (as evidenced by the extensive domain structure seen in this region (*see Supplementary Information* Figure S4)). It is also entirely possible that some permanent polarisation can be induced in the $BaTiO_3$-rich shells on the application of an external field, although little domain structure is currently seen in an unpoled material, in much the same way that some other Bi-containing Pb-free materials display significant piezoelectric response despite having less in the way of obvious large scale domains. It is likely that creating a clearer compositional separation that minimises the volume percentage of the ceramic in a non-polar pseudocubic phase and maximises the amount of ferroelectric material of different compositions both towards the BiFeO$_3$ and BaTiO$_3$ ends of the tie line between the two compositions contributes to the improved long-range ordering under an electric field that results in high remanent polarisation (*see Supplementary Information* Figure S5).

Finally, the surprise in this work is the appearance of the inner core, which is richer in Ba but very low in Ti. In fact, there seems to be a straightforward core-shell segregation between Ti and Fe, where Ti tends towards the shells and Fe towards the cores. The complexity comes from the fact that the Ba and Bi do not always follow this Fe-Ti segregation. The composition seems to be heading in the direction of a perovskite phase of something like BaFeO$_3$ (or more likely, BaFeO$_{2.5}$), but not so far away in composition from the BiFeO$_3$-rich ceramic (containing a little Ti) that is in the outer core. Structurally, however, it is totally distinct from the outer core and consistently has no ½{*ooo*} superlattice spots. This was not just this grain, but similar conclusions were found for another grain containing a Ba-rich inner core. Thus, we again identify a pseudocubic phase. The lattice parameter, whilst not quite as large as that for the shell region, was clearly larger than for most of the outer core. Seeing as it is structurally very similar to the shell, although with a rather different chemistry, it is no surprise that this was not identified in refinements of X-ray diffraction data. Interestingly, we have so far been unsuccessful in discerning any domain structure in this inner core region (see *Supplemental Materials* Figure S4) and it may be non-ferroelectric, or only polarises with field on it, but retains little domain structure at zero field. The question remains as to why this phase even exists at all and why it is pseudocubic. BaFeO$_3$-BiFeO$_3$ solid solutions in the middle of the tie line tend to form tetragonal structures in a centrosymmetric space group (No. 123 - P4/mmm) with a very small c/a ratio, which might explain why our inner core appears pseudocubic and shows no signs of domain structure. This still does not explain fully why it should be favoured to form under these conditions, but it maybe explains why the structure seen is formed when this composition results from the segregation frozen in by quenching. Further work, for example using thermodynamic modelling, would be needed to elucidate why the observed phase segregation is exhibited in this system.

**Conclusion**

The nanoscale structure and chemistry within the core-shell structure of a quenched $Bi(Fe_{0.97}Ti_{0.03})O_3$-$BaTiO_3$ mixed oxide ceramic has been investigated using various STEM techniques. EELS mapping was used to map the Ba, Fe and Ti contents simultaneously, while mapping of Bi content was done separately using the Bi $M_{4,5}$ edge, due to its high energy. To correlate the chemical mapping with the crystallography, SPED was performed along a $<112>_{primitive}$ direction in the same grain using a prototype system with a direct electron detector, and the data was analysed using open-source routines to determine the integrated intensity for superlattice reflections corresponding to antiphase octahedral tilting. The same dataset was analysed to determine the shifts of the diffraction spots and thereby map the lattice parameter with respect to a reference pattern.

The work clearly shows that there are three distinct chemistries present, each having a distinct crystallographic structure associated with it. Firstly, an apparently-pseudocubic shell that is enriched in Ba and Ti (and a relatively large lattice parameter). Secondly, an octahedrally tilted outer core richest in Bi and Fe (showing variation in lattice parameters in different directions, suggesting an inherent domain structure as corroborated by dark field TEM images), fitting well with expectations of an *R3c* $BiFeO_3$-like structure as a major phase in this system. And thirdly and finally, an apparently-pseudocubic inner core phase richer in Ba and very low in Ti (and a lattice parameter similar to the $BaTiO_3$-rich shell), which may indicate a phase somewhere between $BiFeO_3$ and $BaFeO_3$. By comparison with previous literature and by the absence of any discernible domain structure, the inner core phase appears to be non-polar and therefore, unlikely to be involved in the overall ferroelectric properties.

This work clearly demonstrates the power of combining precession electron diffraction and EELS for correlative studies of nanoscale structure and chemistry in complex mixed oxide ceramics, especially when using new electron counting detectors for the SPED. We expect this will be of much wider applicability in a range of different ceramic materials with deliberately inhomogeneous microstructures.

## Acknowledgements

S.J.M. would like to acknowledge the CDT in Photonic Integration and Advanced Data Storage (PIADS) for the funding of his PhD through the EPSRC grant (EP/L015323/1).

We gratefully acknowledge funding from EPSRC which has supported the development of the detector (through "Fast Pixel Detectors: a paradigm shift in STEM imaging" (EP/M009963/1) & Impact Acceleration Accounts (EP/K503903/1 & EP/R511705/1)). Funding from EPSRC (EP/R511705/1) and NanoMEGAS supported integration of the Merlin for EM detector with the ASTAR system. I.C. thanks the National Education of Turkish Republic for financial support throughout his PhD.

# Supplemental Materials

**Exemplar diffraction patterns from <110>$_{primitive}$ and <112>$_{primitive}$ in BiFeO$_3$**

Figure S1 shows schematic calculated diffraction patterns (CrystalMaker / SingleCrystal – CrystalMaker Software Ltd., Oxfordshire, UK) for BiFeO$_3$ (using the structure of Sosnowska *et al.* [1]) for the two crystallographically distinct <110>$_{primitive}$ and four (technically three) crystallographically distinct <112>$_{primitive}$ directions in this structure. For the <110>$_{primitive}$ directions, 50% have no superlattice spots ½{*ooo*} positions (where *o* stands for an odd number), in accordance with the conclusions of Woodward and Reaney[2]. For the <112>$_{primitive}$ directions, all directions have at least some of these superlattice spots, and even if there are systematic absences, these will still see diffraction spots in practice due to double diffraction. Thus, looking for the presence or absence of the ½{*ooo*} in <112>$_{primitive}$ directions is the most robust way in diffraction for determining whether or not antiphase tilting is present.

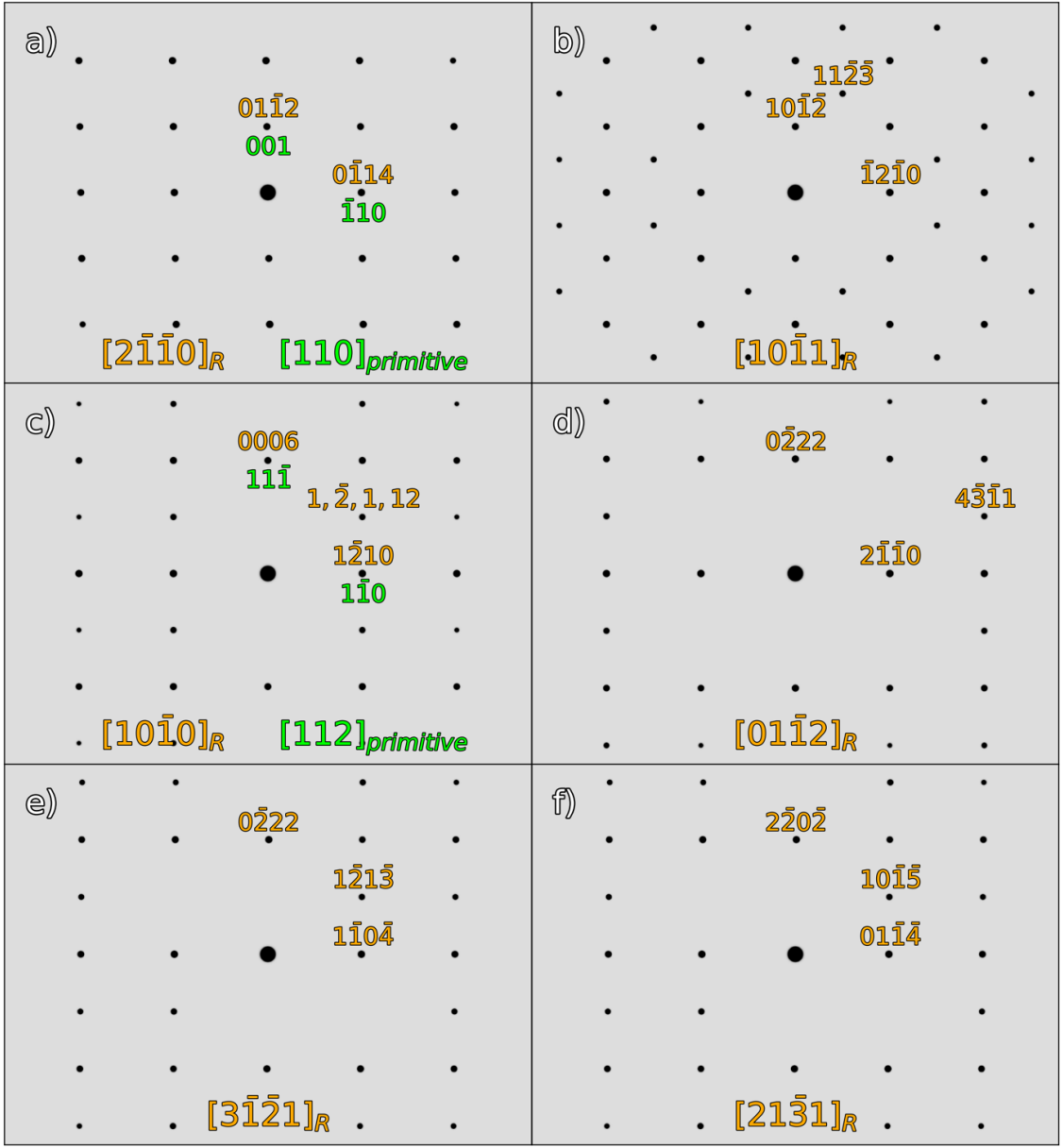

*Fig S1: Schematic diffraction patterns for the two distinct <110>$_{primitive}$ directions and four distinct <112>$_{primitive}$ directions in the BiFeO$_3$. Each of these has 5 more equivalent directions related by symmetry operations (giving 12 directions for <110> and 24 for <112>). Only those <110>$_{primitive}$ directions not perpendicular to the **c**-axis of the rhombohedral give additional spots in positions that would be equivalent to ½{ooo} for the primitive cell. On the other hand, all <112>$_{primitive}$ directions give superlattice spots. In no case is every ½{ooo}$_{primitive}$ position filled, and some are clearly kinematically forbidden, but spots will still appear in practice in these positions due to double diffraction (for which there are plentiful pathways).*

A full list of equivalent directions is shown in Table 1 below. The 4-vector indexing system is used throughout as it most clearly demonstrates symmetry relationships between different vectors and planes in hexagonal systems or rhombohedral systems indexed in a hexagonal setting (as is generally the case for $BiFeO_3$).

| Primitive | [110] | [110] | [112] | [112] | [112] | [112] |
|---|---|---|---|---|---|---|
| Hex 1 | $[10\bar{1}1]$ | $[2\bar{1}\bar{1}0]$ | $[10\bar{1}0]$ | $[1\bar{1}02]$ | $[3\bar{1}\bar{2}1]$ | $[\bar{3}211]$ |
| Hex 2 | $[0\bar{1}11]$ | $[\bar{1}2\bar{1}0]$ | $[0\bar{1}10]$ | $[01\bar{1}2]$ | $[\bar{2}3\bar{1}1]$ | $[1\bar{3}21]$ |
| Hex 3 | $[\bar{1}101]$ | $[\bar{1}\bar{1}20]$ | $[\bar{1}100]$ | $[\bar{1}012]$ | $[\bar{1}\bar{2}31]$ | $[21\bar{3}1]$ |
| Hex 4 | $[\bar{1}01\bar{1}]$ | $[\bar{2}110]$ | $[\bar{1}010]$ | $[\bar{1}10\bar{2}]$ | $[\bar{3}12\bar{1}]$ | $[3\bar{2}\bar{1}\bar{1}]$ |
| Hex 5 | $[01\bar{1}\bar{1}]$ | $[1\bar{2}10]$ | $[01\bar{1}0]$ | $[0\bar{1}1\bar{2}]$ | $[2\bar{3}1\bar{1}]$ | $[\bar{1}32\bar{1}]$ |
| Hex 6 | $[1\bar{1}0\bar{1}]$ | $[11\bar{2}0]$ | $[1\bar{1}00]$ | $[1\bar{1}0\bar{2}]$ | $[123\bar{1}]$ | $[\bar{2}13\bar{1}]$ |

*Table S1: Symmetry equivalent directions in rhombohedral $BiFeO_3$ corresponding to <110>$_{primitive}$ and <112>$_{primitive}$ directions, all expressed in hexagonal 4-vector indices.*

Fig S2 shows the EELS spectra of the three chemically segregated phases of the grain shown in fig 5 and 6 denoted as *shell*, *outer core* and *inner core* as denoted in the main text. This data was obtained in a separate EELS acquisition, in which the energy range of 1812-3848eV with a dispersion of 1eV/channel, so that Bi could be mapped directly. Figure S3 (a) shows the spectra for each region (averaged over 80 pixels) in the energy range containing the Bi M edge. The spectra have been normalised within the range of 2451-2551eV (just before the peak), so as variations in the Bi edge can be discerned between the three regions. (b) shows the absolute counts for each of the spectra after a background subtraction.

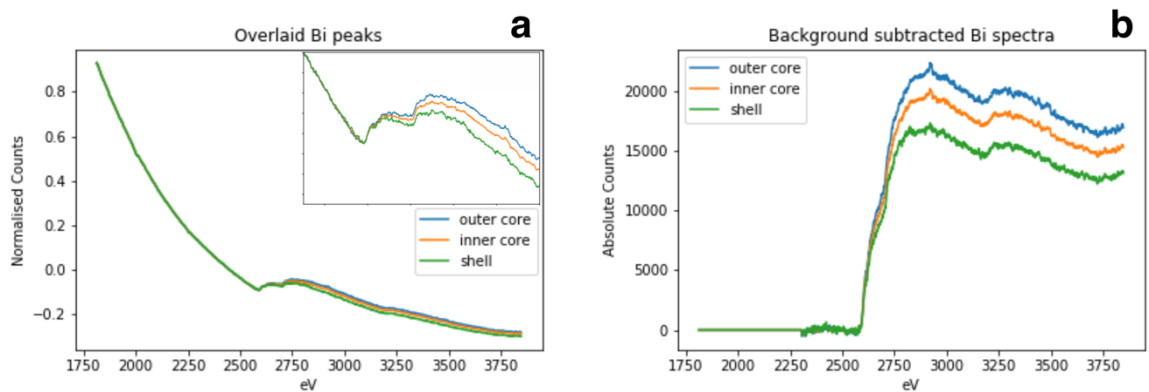

*Fig S2: (a) Normalised spectra for each of the regions showing chemical segregation overlaid, for the EELS acquisition which captured the Bi edge. Inset shows a zoomed in image of the edge region. (b) Absolute counts of each of the spectra after background removal.*

Fig S3 shows an EELS map of the Bi signal obtained of the grain of shown in fig 5 and 6 in the main text. It is clear that the Bi signal correlates with the HAADF image obtained simultaneously, confirming that the intensity variation in the HAADF image can be used as a proxy for a bismuth map in datasets which did not include a prominent bismuth edge within the acquired energy range.

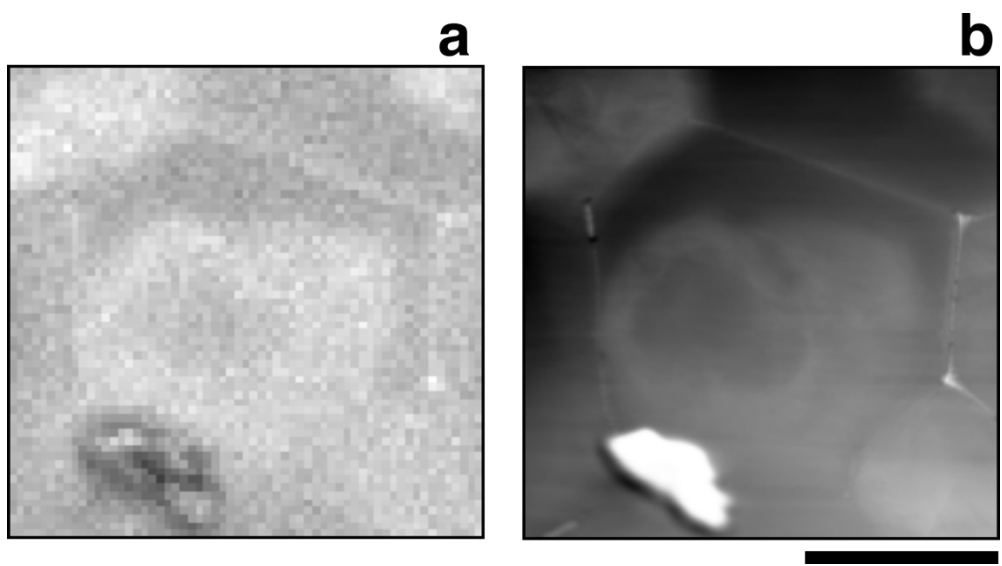

*Fig S3: (a) Bi M signal mapped using the elemental quantification plugin in DM. (b) Simultaneously acquired HAADF signal of the same region. Scale bar represents 500nm.*

Figure S4 shows a TEM dark field image obtained using the $01\bar{1}$ reflection alongside the diffraction pattern to which the grain was orientated, the $[111]_{pc}$ direction. The dark field image clearly shows the presence of a domain structure in the outer core phase ($BiFeO_3$ rich), while the shell and inner core phase show a clear lack of domain structure, which would be expected for a pseudocubic phase.

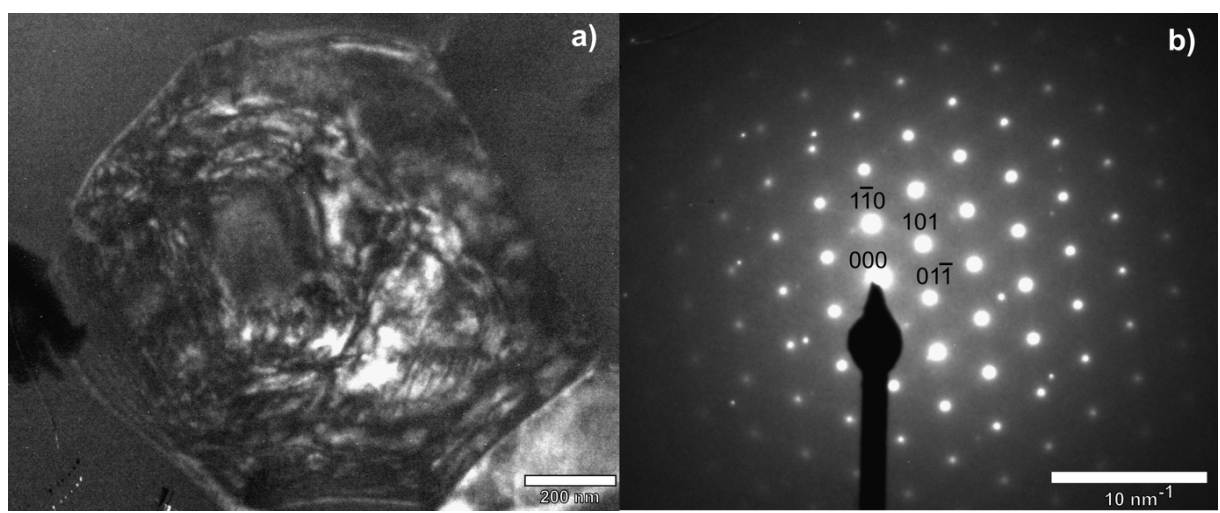

*Fig S4: (a) Dark field image of the grain analysed in the main manuscript (b) Corresponding diffraction pattern to which the grain was aligned. The dark field image was formed using the $01\bar{1}$ reflection.*

Figure S5 shows the polarisation-electric field hysteresis loops for both slow-cooled and quenched ceramics of this composition, clearly showing a high remanent polarisation for the quenched sample with respect to the slow-cooled.

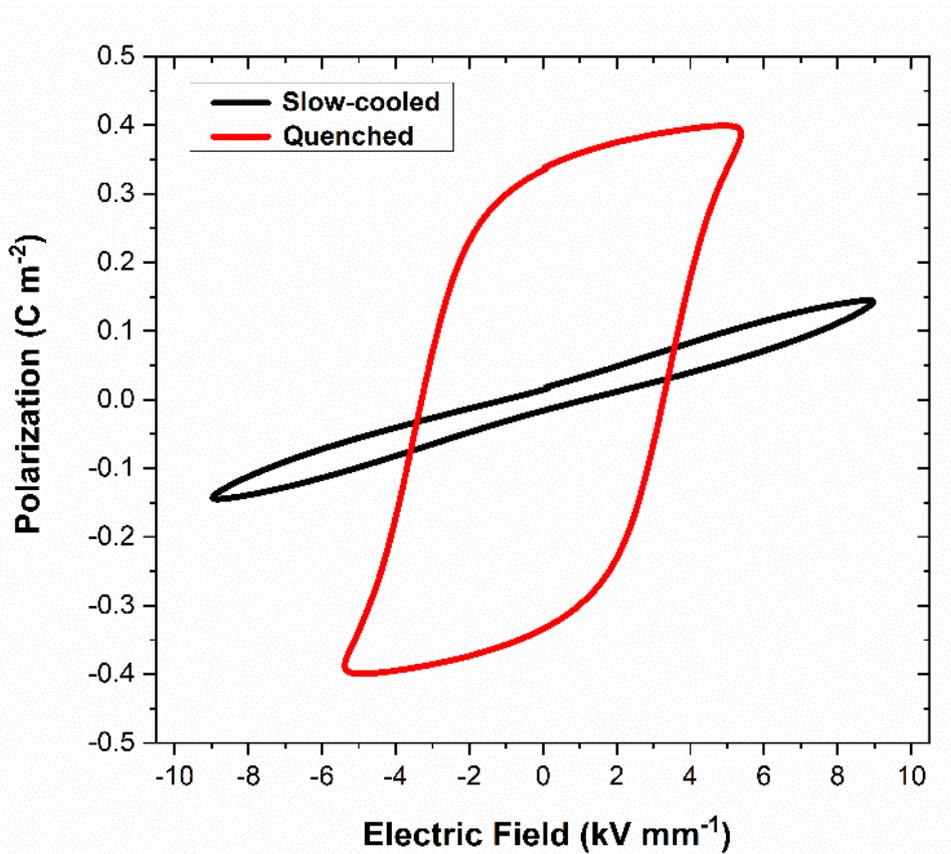

*Fig S5: (a) Comparison of ferroelectric hysteresis loops in slow-cooled and quenched $0.75Bi(Fe_{0.97}Ti_{0.03})O_3 - 0.25BaTiO_3$ ceramics.*